\documentclass[preprint,12pt]{elsarticle}

\usepackage{amssymb}

\newcommand{\icm}{\ensuremath{~\textrm{cm}^{-1}}}
\newcommand{\RFS}{Rb$_{0.8}$Fe$_{1.68}$Se$_2$}
\newcommand{\BKFA}{Ba$_{0.6}$K$_{0.4}$Fe$_{2}$As$_{2}$}

\begin{document}

\begin{frontmatter}

\title{Electron-phonon coupling in the superconducting single crystal Rb$_{0.8}$Fe$_{1.68}$Se$_2$}

\author{B. Xu}
\author{Y. M. Dai\fnref{myfootnote}}
\fntext[myfootnote]{Present address: Condensed Matter Physics and Materials Science Department, Brookhaven National Laboratory, Upton, New York 11973, USA}
\author{J. Han, K. Wang, R. Yang, Y. X. Yang, W. Zhang, H. Xiao}
\author{X. G. Qiu\corref{mycorrespondingauthor}}
\cortext[mycorrespondingauthor]{Corresponding author}
\ead{xgqiu@iphy.ac.cn}

\address{
Beijing National Laboratory for Condensed Matter Physics, Institute of Physics, Chinese Academy of Sciences, P.O. Box 603, Beijing 100190, China}

\begin{abstract}
The optical properties of the superconducting single crystal Rb$_{0.8}$Fe$_{1.68}$Se$_2$ with $T_{c}$ $\simeq$ 31~K have been measured over a wide frequency range in the \emph{ab} plane. We found that the optical conductivity is dominated by a series of infrared-active phonon modes at low-frequency region as well as several other high-frequency bound excitations. The low-frequency optical conductivity has rather low value and shows quite small Drude-like response, indicating low carriers density in this material. Furthermore, the phonon modes increase continuously in frequency with decreasing temperature; specifically, the phonon mode around 200\icm\ shows an enhanced asymmetry effect at low temperatures, suggesting an increasing electron-phonon coupling in this system.
\end{abstract}


\end{frontmatter}


\section{Introduction}

The discovery of superconductivity in the ternary iron selenides A$_{y}$Fe$_{2-x}$Se$_2$ (A = K, Rb, Cs, ...)~\cite{Guo2010,Wang2011a,Krzton2011,Fang2011} has triggered great interest not only because of its relatively high value for $T_c$ (over 30~K), but also for its unique properties absent in other iron-based superconductors. First, in this system the superconductivity is found to be induced from an antiferromagnetically insulating phase~\cite{Fang2011}, different from other iron-based superconductors where the superconductivity is evolved from a spin-density-wave type metal~\cite{Dong2008,Cruz2008,Singh2009}. Further, the hole pockets at the center of the Brillouin zone are absent, leaving just the electron pockets at the corner of the zone~\cite{Nekrasov2011,Yan2010,Zhang2011,Qian2011}, which suggests that the inter-pocket scattering between the hole and electron pockets is not an essential ingredient for superconductivity in this system. Moreover, Fe vacancies are found to be present and form the $\sqrt{5} \times \sqrt{5} \times 1$ $I4/m$ ordered pattern in the lattice~\cite{Ye2011,Zhang2012,Bao2011,Wang2011}. In addition, evidence for phase separation and coexistence of magnetism and superconductivity has also been observed~\cite{Bao2011,Wang2011,Torchetti2011,Zhao2012,Texier2012,Friemel2012,Ricci2011,Charnukha2012a,Yuan2012,Homes2012}.

The superconductivity in iron-based superconductors is believed to occur beyond the conventional electron-phonon coupling mechanism~\cite{Boeri2008,Noffsinger2009}, and the pairing mechanism is proposed to be mediated by exchange of the antiferromagnetic spin fluctuations~\cite{Mazin2008,Kuroki2008}. In this newly discovered iron selenides system, a number of Raman-active phonon modes~\cite{Zhang2012,Ignatov2012} as well as phonon anomalies associated with a rather specific type of electron-phonon coupling~\cite{Zhang2012} have been observed in Raman-scattering studies. Moreover, the measurements of optical conductivity have also identified a series of infrared-active phonon modes in the far infrared region as well as a number of high-frequency bound excitations~\cite{Yuan2012,Homes2012,Chen2011,Charnukha2012}. Additionally, nuclear magnetic resonance (NMR) experiments suggest that spin fluctuations are weak~\cite{Torchetti2011,Yu2011,Kotegawa2011}. All the above results give evidence for the possible role played by electron-phonon coupling to the appearance of superconductivity in this system. In this work, we performed optical spectroscopy measurements on the superconducting single crystal of \RFS\ with $T_{c}$ $\simeq$ 31~K. Our results show that the \RFS\ presents similar optical conductivity features as the previous reports in K$_{y}$Fe$_{2-x}$Se$_2$~\cite{Yuan2012,Homes2012,Chen2011}. Moreover, we found that the phonon mode around 200\icm\ displays remarkable asymmetry effect at low temperatures. A detailed Fano line shape analysis suggests a strong coupling between lattice and electronic background.

\section{Experiments}

High quality \RFS\ ($T_c$ $\simeq$ 31~K) single crystals were grown by the Bridgeman method and the composition of the samples were determined by the ICP analysis method~\cite{Li2011}. The \emph{ab}-plane reflectivity $R(\omega)$ was measured at a near-normal angle of incidence on BOMEN DA8 FT-IR spectrometers. An \emph{in situ} gold overfilling technique~\cite{Homes1993} was used to obtain the absolute reflectivity of the sample. Data from 40 to $12\,000\icm$ were collected at different temperatures from 5~K to 300~K on freshly cleaved surfaces on an ARS-Helitran crysostat. We extended the reflectivity to $50\,000\icm$ at room temperature with an AvaSpec-2048 $\times$ 14 optical fiber spectrometer. The real part of the optical conductivity $\sigma_1(\omega)$ was determined from the reflectivity through Kramers-Kronig analysis~\cite{Dressel2002}.

\section{Results and discussion}

\begin{figure}
\begin{center}
\includegraphics[width=100mm]{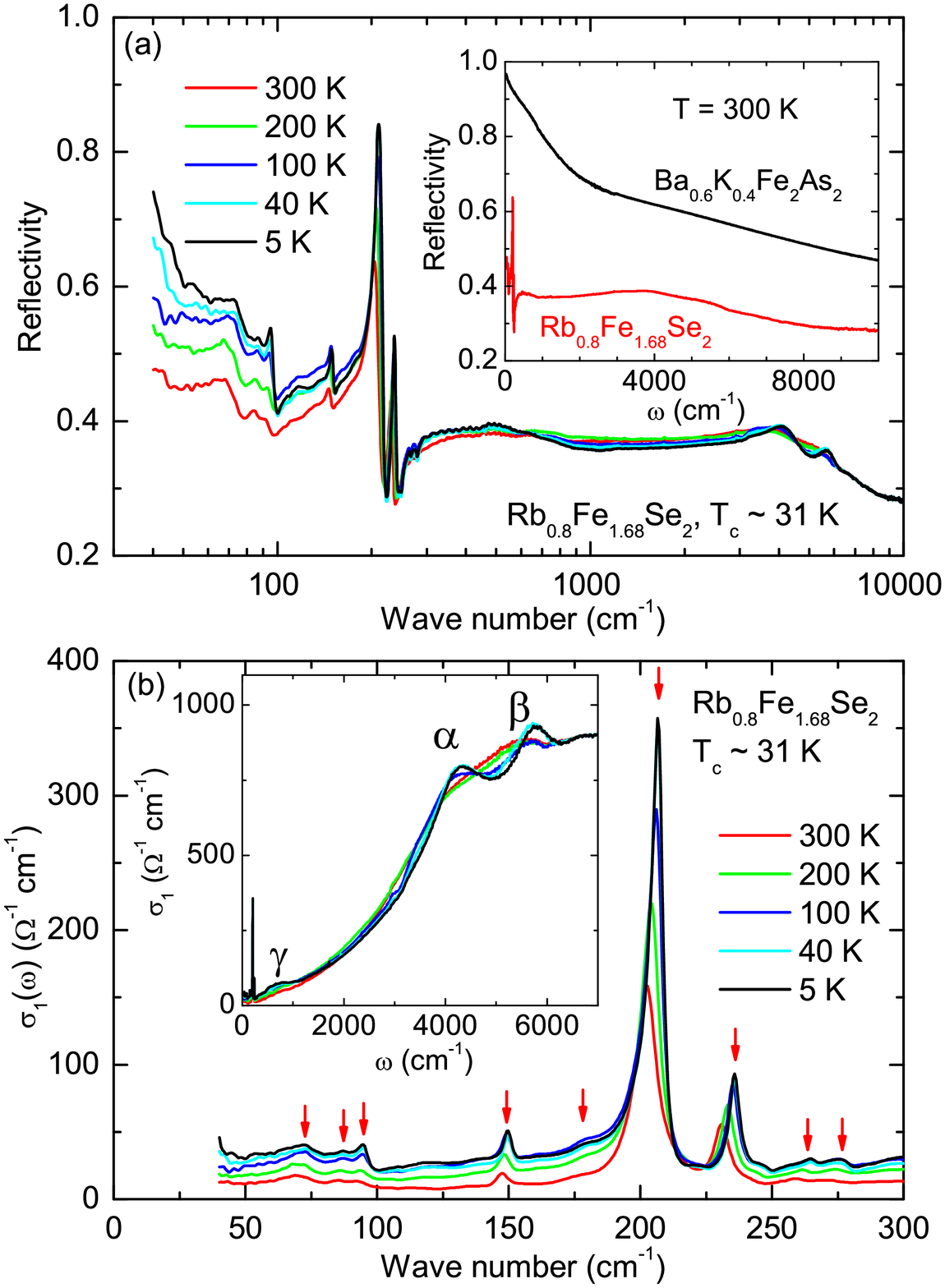}
\caption{\label{fig.1} (color online) (a) The reflectivity over a wide frequency range for \RFS\ at selected temperatures. Inset: a comparison for the reflectivity between \RFS\ and \BKFA\ at 300~K. (b) The real part of the optical conductivity $\sigma_1(\omega)$ in the low-frequency region for \RFS\ at selected temperatures. Inset: the conductivity $\sigma_1(\omega)$ spectrum over a wide frequency region. Three prominent bound excitation features are located at 700, 4300, and 5700\icm, and are labeled as $\gamma$, $\alpha$, and $\beta$, respectively.}
\end{center}
\end{figure}

Figure~\ref{fig.1}(a) shows the temperature dependence of the reflectivity $R(\omega)$ for \RFS\ over broad frequencies up to 10\,000\icm. Over all, the reflectivity has a rather low value around 0.4 reflects the characteristic of a poor metal. As a comparison, we plot the reflectivity for \RFS\ and \BKFA\ at 300~K in the inset of Fig.~\ref{fig.1}(a). The reflectivity for \BKFA\ is about twice higher than that for \RFS. Moreover, the reflectance spectrum for \RFS\ is dominated by a series of sharp features associated with the infrared-active vibrations in the low-frequency region as well as other high-energy electronic state features.

Figure~\ref{fig.1}(b) displays the temperature dependence of the conductivity $\sigma_1(\omega)$ for \RFS\ in the low-frequency region. At room temperature $\sigma_1(\omega)$ is dominated by a series of infrared-active phonon modes (indicated by the red arrows in Fig.~1(b)) superposed on a flat and incoherent electronic background. The large number of the low-frequency infrared-active phonon modes is beyond the observation in BaFe$_2$As$_2$ ($I4/mmm$), where only two in-plane infrared-active phonon modes are observed~\cite{Akrap2009}. As the temperature is reduced, the phonon modes narrow and increase slightly in frequency, meanwhile, the electronic background increases slightly and a small Drude-like response emerges. In addition to the low-frequency vibrational features, there are several broad bound excitation features observed at high frequency, as shown in the inset of Fig.~\ref{fig.1}(b). The three prominent features are located at 700, 4300, and 5700\icm, which have also been observed in the K- and Rb-doped iron selenides in previous optical studies~\cite{Yuan2012,Homes2012,Chen2011,Charnukha2012}, and are labeled as $\gamma$, $\alpha$, and $\beta$, respectively. The high-frequency bound excitation features were thought to be highly related to the Fe vacancies and their orderings in this system~\cite{Chen2011}, where the Fe vacancies form a $\sqrt{5} \times \sqrt{5} \times 1$ superlattice pattern~\cite{Ye2011,Zhang2012,Bao2011,Wang2011}.

The optical conductivity can be described by the Drude-Lorentz model:
\begin{equation}
\label{DrudeLorentz}
\sigma_{1}(\omega)=\frac{2\pi}{Z_{0}}\biggl[\sum_{k}\frac{\Omega^{2}_{p,j}}{\omega^{2}\tau_j + \frac{1}{\tau_j}} + \sum_{k}\frac{\gamma_{k}\omega^{2}S^{2}_{k}}{(\omega^{2}_{0,k} - \omega^{2})^{2} + \gamma^{2}_{k}\omega^{2}}\biggr],
\end{equation}
where $Z_{0}$ is the vacuum impedance. The first term describes a sum of free-carrier Drude responses, each characterized by a plasma frequency $\Omega_{p,j}$ and a scattering rate $1/\tau_j$. The square of plasma frequency $\Omega^2_{p}$ is proportional to $n/m$, where $n$ and $m$ refer to the density and effective mass of conduction electron. The second term is a sum of Lorentz oscillators, each having a resonance frequency $\omega_{0,k}$, a line width $\gamma_k$ and an oscillator strength $S_k$.

\begin{figure}
\begin{center}
\includegraphics[width=100mm]{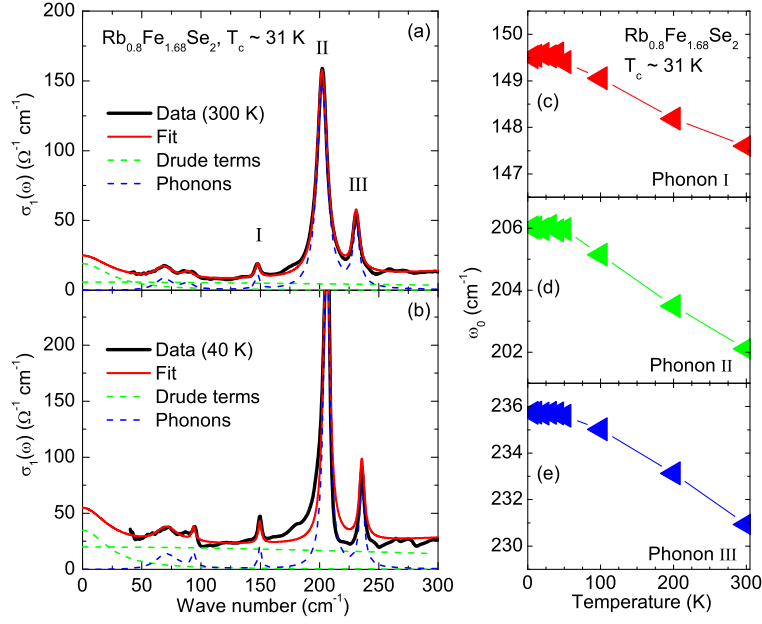}
\caption{\label{fig.2}  (color online) The Drude-Lorentz fit to the optical conductivity $\sigma_1(\omega)$ for \RFS\ at (a) 300~K and (b) 40~K. Temperature dependence of the frequency $\omega_0$ for the three most prominent phonon modes observed around (c) 230\icm, (d) 200\icm, and (e) 150\icm, respectively.}
\end{center}
\end{figure}

Figure~\ref{fig.2} shows the results of the Drude-Lorentz fit to the conductivity at 300~K [Fig.~\ref{fig.2}(a)] and 40~K [Fig.~\ref{fig.2}(b)]. We use five Lorentz oscillators to reproduce the five major phonon modes and one Lorentz components to describe the interband transition near 700\icm. In addition to these Lorentz terms, we have to use a narrow and a broad Drude terms to reproduce the low-frequency free carrier contributions. As can be seen, the model reasonably reproduces the $\sigma_{1}(\omega)$ spectrum. The plasma frequency of the Drude terms is $\Omega_{p,1} \simeq$ 178\icm\ and $\Omega_{p,2} \simeq$ 368\icm\ at 300~K, $\Omega_{p,1} \simeq$ 230\icm\ and $\Omega_{p,2} \simeq$ 742\icm\ at 40~K. The total plasma frequency is given by $\Omega_{pD} = \sqrt{\Omega^2_{p,1} + \Omega^2_{p,2}}$. Thus the value of $\Omega_{pD}$ is $\sim$ 408\icm\ at 300~K and $\sim$ 776\icm\ at 40~K, which is about one order smaller than those of other iron-based superconductors~\cite{Hu2008,Tu2010,Dai2013}. The small value of $\Omega_{pD}$ is consistent with the low carrier density in this material.

The vibrational parameters for the three most prominent phonon modes in \RFS\ observed around 230\icm, 200\icm, and 150\icm\  at 300~K and 5~K are summarized in Table~\ref{tab.1}. The detailed temperature dependence of the phonon frequency for the three phonon modes are shown in Fig.~2(c-e). As we can see, the phonon frequency increases continuously with decreasing temperature. Additionally, as shown in Fig.~\ref{fig.2}(b), we notice that there exists a strong asymmetric effect for the phonon mode around 200\icm\ at low temperatures, while the Lorentzian oscillator was insufficient to accurately describe this line shape. The asymmetric line shape is general, and coupling of a lattice mode to a continuum of charge or spin excitations often produces such an asymmetric line shape~\cite{Fano1961,Dowty1987}.

The Fano line shape analysis can help us to understand the underlying coupling. Following the Fano theory~\cite{Fano1961}, the optical conductivity $\sigma_{1}(\omega)$ of such coupled system can described by the formula~\cite{Kuzmenko2009}:
\begin{equation}
\label{Fano}
\sigma_{1}(\omega)=\frac{2\pi}{Z_0} \frac{S^2}{\gamma}
\frac{q^2 +\frac{4q (\omega - \omega_0)}{\gamma} -1}{q^2 (1 + \frac{4(\omega - \omega_0)^2}{\gamma^2})}.
\end{equation}
The above Fano line shape description is a modified form of the standard Lorentz oscillator. The asymmetry is described by a dimensionless parameter $q$, where the symmetric Lorentz line shape is completely recovered in the limit $1/q \rightarrow 0$. The parameter $q$ is inversely related to the strength of the interaction. The sign of $q$ determines the energy range that the phonon is coupled. For example, in the case of $q < 0$, a line shape that dips on the high frequency side of the phonon frequency $\omega_0$, indicates the phonon is interacting with a continuum at lower energies.

\begin{figure}
\begin{center}
\includegraphics[width=100mm]{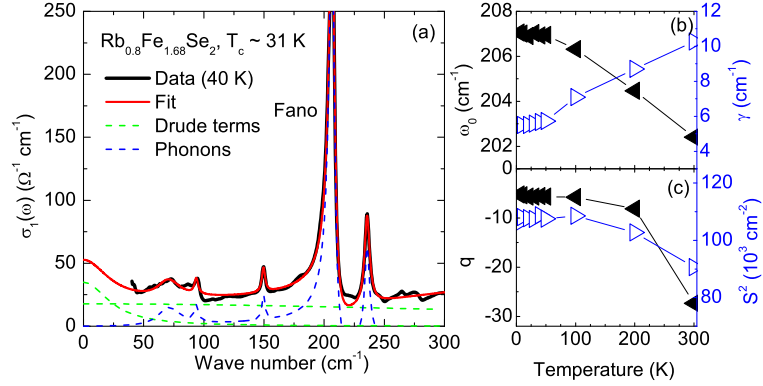}
\caption{\label{fig.3} (color online) (a) The alternative Fano Fit for the phonon mode around 200\icm\ at 40~K in \RFS. Temperature dependence of (a) the frequency $\omega_0$ (black triangle) and line width $\gamma$ (blue open triangle), as well as (c) the Fano parameter $q$ (black triangle) and the intensity $S^2$ (blue open triangle) of the phonon mode around 200\icm.}
\end{center}
\end{figure}

As shown in Fig.~\ref{fig.3}(a), the sharp phonon mode around 200\icm\ can be well reproduced by replacing the Lorentz oscillator with the Fano oscillator. The Fano fit parameters at 300~K and 5~K are shown in Table~\ref{tab.1}. We can see that this phonon mode shows a strong asymmetry with $q \simeq -5.3$ at 5~K, and presents a weak asymmetry at room temperature ($q \simeq -27.3$) that tends to a Lorentz line shape. The negative sign of $q$ suggests that this phonon mode interacts with a continuum at lower energies. The temperature dependence of the Fano parameters for this phonon mode is shown in Figs.~\ref{fig.3}(b)--\ref{fig.3}(c). From Fig.~\ref{fig.3}(b), one can see that $\omega_0$ increases continuously and the line width $\gamma$ narrows almost linearly with decreasing temperature. Note that due to the Fano line shape, $\omega_0$ is larger than the position of the Lorentz oscillator fit, the difference being dependent on the asymmetry parameter $q$, $\Delta \omega_0 \simeq $ 0.3\icm\ at 300~K and $\Delta \omega_0 \simeq $ 1.1\icm\ at 5~K. The black triangle in Fig.~\ref{fig.3}(c) shows the temperature dependence of the asymmetry parameter $q$. As we can see, the asymmetry effect increases with decreasing temperature and becomes saturated at low temperatures. Meanwhile, as shown by the blue open triangle in Fig.~\ref{fig.3}(c), we can also see a corresponding increase in the lattice mode intensity.

The enhanced asymmetry effect at low temperatures implies an increased spin-lattice or electron-lattice coupling. Nevertheless, the NMR experiments have evidenced that the spin lattice relaxation rate $1/T_1$ in this system is decreased at low temperatures~\cite{Torchetti2011,Yu2011,Kotegawa2011}. Thus, this brings up the possibility of changes to the electronic states in this system. Indeed, our results as well as the previous optical results~\cite{Yuan2012,Homes2012,Charnukha2012} all suggest that the optical conductivity in this system is almost incoherent at room temperature and becomes more coherent at low temperatures. Therefore, the increase in asymmetry can be understood by considering that the coherent electronic states are continuously increased as the temperature is reduced. Such increase of the coherent quasiparticles results in the enhanced coupling of the lattice mode to the electronic states.

\section{Conclusion}

In summary, we performed optical measurements on the superconducting single crystal of \RFS. We found that the optical conductivity is dominated by a series of sharp features associated with the infrared-active phonon modes in the low-frequency region as well as several high-frequency bound excitations. As the temperature is reduced, the phonon modes increase continuously in frequency and a Drude-like low-frequency response emerges, but the fitted value $\Omega_{pD}$ is quite small, indicating low carriers density in this material. Moreover, the phonon mode around 200\icm\ shows a strong asymmetry effect and a corresponding increase in the phonon mode intensity at low temperatures due to the increased coherent electronic states, suggesting a strong electron-phonon coupling in this system. This electron-phonon coupling may play a role in the the appearance of superconductivity in this iron selenides system.

\section*{Acknowledgments}
We thank C. H. Li, and H. H. Wen for providing \RFS\ crystals. This work was supported by MOST (973 Projects No. 2012CB821403, 2011CBA00107, 2012CB921302), and NSFC (Grants No. 11374345, 11104335 and 91121004).

\section*{References}


\begin{table}
\caption{\label{tab.1} The vibrational parameters for oscillator fits to the three most prominent phonon modes in \RFS\ observed around 230\icm, 200\icm, and 150\icm\ at 300~K and 5~K, where $\omega_{0}$, $S$, and $\gamma$ are the oscillator frequency, strength, and width, respectively. Units for $\omega_0$, $S$, and $\gamma$ are in cm$^{-1}$. $q$ is the Fano asymmetry parameter.}
\begin{center}
\begin{tabular}{l|cccc}
\hline
Phonon & $\omega_{0}$ & $S$ & $\gamma$ & $q$ \\
\hline
I (300K) & 147.6 & 52.4 & 4.5 & -- -- \\
I (5K)& 149.5 & 63.6 & 3.1 & -- -- \\
II (300K)& 202.1 & 300.2 & 10.2 & -- -- \\
II (300K, Fano) & 202.4 & 301.0 & 10.3 & --27.3 \\
II (5K)& 206.0 & 330.6 & 5.6 & -- -- \\
II (5K, Fano) & 207.1 & 325.7 & 5.5 & --5.3 \\
III (300K) & 230.9 & 129.7 & 6.6 & -- -- \\
III (5K) & 235.7 & 147.8 & 5.0 & -- -- \\
\hline
\end{tabular}
\end{center}
\end{table}

\end{document}